# Machine Learning-Assisted Nano-imaging and Spectroscopy of Phase coexistence in a Wide-Bandgap Semiconductor

*Alyssa Bragg[1], Fengdeng Liu[2], Zhifei Yang[1,2], Nitzan Hirshberg[1], Madison Garber[1], Brayden Lukaskawcez[1], Liam Thompson[1], Shane MacDonald[1], Hayden Binger[1,†], Devon Uram[1], Ashley Bucsek[3], Bharat Jalan[2], Alexander McLeod[1]*

[1]School of Physics and Astronomy, University of Minnesota – Twin Cities, Minneapolis, Minnesota 55455, United States

[2]Department of Chemical Engineering and Materials Science, University of Minnesota – Twin Cities, Minneapolis, Minnesota 55455, United States

[3]Department of Mechanical Engineering, University of Michigan, Ann Arbor, Michigan 48109, United States

**ABSTRACT.** Wide bandgap semiconductors with high room temperature mobilities are promising materials for high-power electronics[1–3]. Stannate films provide wide bandgaps and optical transparency, although electron-phonon scattering can limit mobilities[4]. In $SrSnO_3$, epitaxial strain engineering stabilizes a high-mobility tetragonal phase at room temperature, resulting in a threefold increase in electron mobility among doped films[1]. However, strain relaxation in thicker films leads to nanotextured coexistence of tetragonal and orthorhombic phases with unclear implications for optoelectronic performance[5,3]. The observed nanoscale phase coexistence[5] demands nano-spectroscopy to supply spatial resolution beyond conventional, diffraction-limited microscopy. With nano-infrared spectroscopy, we provide a comprehensive analysis of phase coexistence in SrSnO₃ over a broad energy range, distinguishing inhomogeneous phonon and plasma responses arising from structural and electronic domains. We establish Nanoscale Imaging and Spectroscopy with Machine-learning Assistance (NISMA) to map nanotextured phases and quantify their distinct optical responses through a robust quantitative analysis, which can be applied to a broad array of complex oxide materials.

**TEXT.** Wide bandgap semiconductors have high dielectric breakdown strength and are transparent in the visible spectrum, making them of interest for application in high-power optoelectronic devices[6,7]. Doped $BaSnO_3$ (BSO), with a bandgap over 3 eV, has demonstrated high room temperature mobilities in bulk, however, lattice mismatch with substrates makes growing high-mobility films challenging[8–13]. $SrSnO_3$ (SSO) has a wider bandgap of 4.1 eV[14–16], and while it has lower mobility, its smaller lattice parameters allow coherent growth on substrates and provide strain engineering as an avenue to enhance mobility[1,3]. By compressive strain, the high-temperature tetragonal phase can be stabilized at room temperature, compared to 1062 K in bulk[17], resulting in a threefold enhancement in mobility compared to the room temperature orthorhombic phase[1] in doped films. While calculations predict the orthorhombic phase has a 30% greater electron effective mass over the tetragonal phase, this alone does not explain enhanced mobility in tetragonal films, calling for deeper investigation into electronic properties of these phases[1].

Strain and doping serve as parameters to tune structural phases of SSO thin films. Epitaxially strained films grown on $GdScO_3$ (GSO) below a critical thickness are entirely tetragonal. When thickness is increased, the orthorhombic phase emerges and coexists with the tetragonal phase through strain relaxation[3]. La doping through Sr substitution increases the orthorhombic-to-tetragonal transition temperature and introduces defects, increasing the relative volume of tetragonal to orthorhombic phases at room temperature[5]. In SSO films exhibiting tetragonal-orthorhombic phase coexistence, the ability to control high-mobility properties through strain and doping motivates our investigation.

In epitaxially strained thin films of La-doped SSO, nano-optical imaging has revealed that the tetragonal-to-orthorhombic phase transformation drives spontaneous formation of periodic nanostructures[5]. To understand the high-mobility properties of these films, a robust investigation of the nanostructure is required with measurements capable of high spatial resolution. Uniquely suited to this task is near-field optical microscopy, in which light is focused onto an atomic force microscope probe in contact with a sample surface and drives a confined near-field optical interaction[18,19]. Light scattered from the probe is modulated by this interaction, encoding information about the sample's permittivity. The spatial resolution achievable is determined only by the diameter of the tip apex[18,19], as small as 10 nm, in contrast to conventional microscopy, in which spatial resolution is limited by the wavelength of light.

The measured near-field response is most simply related to the material's structural and electronic properties through the Drude-Lorentz model[20,21] (more details given in *Supplementary Information,* or *SI*). The electronic response of a doped semiconductor is dominated by the plasma oscillation at the screened plasma frequency $\omega_{p,sc} = \sqrt{\frac{4\pi n e^2}{m^* \epsilon_\infty}}$, where $n$ is the carrier concentration, $m^*$ is the effective mass of carriers and $\epsilon_\infty$ is the permittivity in the high frequency limit. Lorentz oscillators approximate the response of optical phonons (see *SI*). The permittivity given by the Drude-Lorentz model[22] determines the energy dependent coefficients of reflectance $R$ and absorption $\alpha$ of light from the SSO surface. Fig. 1h presents hypothetical spectra arising from structural phases with distinct optical phonon energies $\omega_{TO}$, such that structural domains are distinguishable by reflectance when probed near $\omega \sim \omega_{TO}$. When dopants are introduced, the plasma resonance also contributes to reflectance, as hypothesized in Fig. 1i for the case where $\omega_{p,sc} > \omega_{TO}$ (more details given in *SI*).

Previous near-field microscopy of La-doped SSO films revealed periodically nanostructured dielectric properties, ascribed to non-uniform carrier mobility among orthorhombic and tetragonal domains[5]. Film thickness and dopant concentration tuned the relative volume fraction of orthorhombic and tetragonal phases, providing an avenue to control dielectric properties and geometry of the dielectric meta-surface, which might enable exotic optical responses[5,23]. While these former results shed light on the formation of tunable nanotextured domains in La-doped SSO, nano-resolved imaging across a wide frequency range in La-doped SSO is needed to confidently assign electronic and structural phases to domain contrasts in imaging. Further, to clarify their role in establishing bulk electronic and optical properties, it is imperative to determine the Drude-Lorentz parameters that describe individual domains. To supply these insights, we apply a comprehensive nano-spectroscopy method with high spatial resolution and broadly tunable access to infrared energies relevant to doped wide-bandgap semiconductors.

We introduce a novel technique to characterize and map coexisting phases in materials based on near-field images collected over a broad spectral range, named Nanoscale Imaging and Spectroscopy with Machine-learning Assistance (NISMA). Near-field images are collected at many energies of interest, informed by area-averaged spectroscopy of the region under study. Using near-field microscopy, we measure both the amplitude and phase of light scattered from the interacting probe-sample system, which we identify as the coefficients of reflectance ($R$) and absorption ($\alpha$)[19] (further details provided in *SI*). We combine reflectance and absorption images acquired from the same sample area collected at 13 distinct energies over a span of 709 cm$^{-1}$ to 1667 cm$^{-1}$ enabled by energy-tunable illumination from an ultrafast laser system (details in *SI*). The nanoscale spectro-microscopy data is understood as a data cube over the sample area, aligned pixel-by-pixel at a spatial resolution of at least 50 nm[24] (described in *SI* ).

We employed principal component analysis (PCA) to reduce dimensionality of the nano-resolved spectro-microscopy data while preserving spectral variation of the real-space images[25] (details in *SI*). The spectral characteristics of each imaged pixel are then sorted into clusters (structural or electronic "phases") by unsupervised machine learning using a Gaussian mixture model (GMM)[26]. Cluster membership allows us to 1) segment the imaged region according to local optical properties, 2) locate four distinct phases with quantified statistical certainty, and 3) classify phases by their Drude-Lorentz parameters.

We use NISMA to uncover nanoscale inhomogeneity in two 72 nm SSO films on GdScO$_3$ (110) substrates: one with a La dopant concentration of $1.17 \times 10^{20}$ cm$^{-3}$ and one without doping, grown by molecular beam epitaxy (details in *SI*)[1–5]. Both films contain coexisting tetragonal and orthorhombic phases as confirmed by X-ray diffractometry (details in *SI*), which reveals the undoped and La-doped films as majority tetragonal and orthorhombic, respectively. For initial characterization, we perform far-field optical microscopy (sketched in Fig. 1a) using a home-built laser-based infrared microscope (details in *SI*) to acquire reflectance images of both films (Figs. 1c-d) at a laser illumination energy of 709 cm$^{-1}$, highlighting contrasts due to the difference in optical phonon energies between the two structural phases. In the undoped film (Fig. 1c), the striped ordering of structural phases can be resolved. However, in the La-doped film (Fig. 1d), no coherent patterns are observable. From the far-field results, we conclude conventional microscopy is insufficiently resolved to probe phase coexistence in the La-doped film. Near-field imaging with nanoscale resolution (Fig. 1e) reveals the striped pattern in undoped SSO, as seen in far-field imaging. However, for the La-doped film, near-field imaging (Fig. 1f) reveals a nanostructure that was not resolvable in far-field imaging, with phase coexistence occurring at much smaller length scales than in the undoped sample.

Near-field reflectance and absorption spectra acquired from the two structural phases (identified in Fig. 1e) in the undoped film are presented in Fig. 1g. The orthorhombic phase has a lower frequency phonon, as indicated by the peaks in reflectance and absorption spectra. Fit Lorentz parameters for the tetragonal and orthorhombic phonons are given in *SI*. Since these phonons result in differential near-field contrast in the energy range 660-740 cm$^{-1}$, imaging at these frequencies distinguishes orthorhombic from tetragonal domains.

While structural domains can be inferred from images, determining what material properties give rise to infrared contrasts requires spectroscopy. Near-field reflectance and absorption spectra shown in Fig. 2b were acquired by nanoFTIR spatially averaged over a 20x20 square microns area of phase coexistence, normalized to the spectrum from an adjacent gold patch deposited on the sample surface (Fig. 2a). Shading in Fig. 2b highlights frequency regimes of three features modelled by the fit (dashed; see *SI*): a phonon response at $\omega_{TO} = 660$ cm$^{-1}$ (Region I), an electronic Lorentzian term at $\omega_{el,1} = 990$ cm$^{-1}$ (Region II), and a Drude response at $\omega_{p,sc} = 1650$ cm$^{-1}$ (Region III). A possible explanation for the electronic Lorentzian term at $\omega_{el,1}$ is a polaron response in the film[27–29], however we do not explore its origin in detail. The fit also features a second electronic Lorentzian term at $\omega_{el,2} = 1930$ cm$^{-1}$, beyond the range relevant to our analysis. This may arise from frequency dependence of the effective mass and scattering rate not captured by our simple Drude model, such as captured by the extended Drude model[30,31].

To spatially resolve local variations in spectral response presented by Fig. 2b, we collected reflectance and absorption images using pseudoheterodyne detection (details in *SI*) of the sample region at 13 distinct energies, denoted at vertical dashed energies in Fig. 2b. Selected images are shown in Fig. 2c. The absorption maps in Region I indicate structural domains, however, in Regions II and III, there is contrast within structural domains, which grows finer with increasing energy. To extract characteristic length scales of periodicity as a function of energy, we computed autocorrelation functions for the absorption images (Fig. 3a); these represent the similarity between an image and a displaced version of itself. Thus, the characteristic length of periodic structures in an image is quantified by the peak-to-peak distance in its autocorrelation. For each image, we take linecuts through the central autocorrelation peak (Fig. 3b) to extract length scales of characteristic periodicity[24], which we label by $\zeta$ or $\xi$, plotted as a function of frequency in Fig. 3c. Whereas an energy-constant length scale $\xi \sim 8$ microns is resolvable at all energies, a smaller length scale $\zeta \sim 2.5$ microns dominates at energies across Regions II and III. If we associate $\xi$ with the periodicity of structural domains most resolvable across Region I, $\zeta$ reports a three-fold finer *intra-domain* contrast associated with the inhomogeneous electronic response. In Fig. 3d we plot spatial variance in infrared absorption as a function of energy, which decreases from Region I to Regions II and III, indicating that structural domains explain most variance in absorption, whereas that associated with $\zeta$ is subdominant. The continued decrease in variance from Region II to Region III may indicate an internal structure not explored in this work.

Such energy-dependent length scales of absorption contrast in La-doped SSO could originate from plasmon-polaritons, as reported in other semiconducting oxides[32]. To determine if length scales of absorption contrast in the 72 nm La-doped SSO are consistent with plasmons, we investigate the plasmon wavelength in a structurally uniform film. We grew a 19 nm La-doped

SSO film with carrier concentration $n = 9.75 \times 10^{19}$ cm$^{-3}$, comparable to the 72 nm La-doped SSO with $n = 1.17 \times 10^{20}$ cm$^{-3}$. This thinner film is uniformly tetragonal (see *SI*) and etched to provide a boundary to launch plasmons, allowing detection through plasmon interferometry (see *SI*). The absorption image taken around this boundary (Fig. 4a) demonstrates variation in absorption as a function of distance from the etched boundary, as well as around a scratch on the sample surface. Absorption linecuts acquired while scanning across these features and incrementally tuning laser energy (Fig. 4b) reveal that the distances between absorption extrema vary as a function of energy, consistent with a plasmon-polariton with wavelength λ$_p$ decreasing with increasing frequency[33,34]. Then, from the reflectance and absorption relative to the GSO substrate, we extract the SSO permittivity $\varepsilon$ as a function of energy for the 19 nm La-doped SSO (Fig. 4c) and confirm the prediction that $\lambda_p \propto 1 - Re(\varepsilon)$, using known optical constants for the GSO[35] (see *SI*). We fit this permittivity to the Drude-Lorentz model and use this to predict the plasmon dispersion (see *SI*). We find that the plasmon momenta, $q_p = 2\pi/\lambda_p$, agree with the modelled dispersion (Fig. 4d), establishing a self-consistent quantification of plasmons in La-doped tetragonal SSO. From our analysis of plasmons in 19 nm SSO, we now compare plasmon wavelength as a function of frequency to length scales of periodicity ζ observed in the 72 nm La-doped SSO (purple symbols in Fig. 3c). For this evaluation, the plasmon wavelength measured from the 19 nm film is rescaled by a factor of 72/19 to account for the difference in film thickness (see *SI*). At all measured frequencies, ζ in the 72 nm La-doped SSO appears larger than (and unrelated to) the plasmon wavelength expected for this film. We conclude that the periodic absorption length scales in the 72 nm La-doped SSO are inconsistent with solely plasmon interference; thus, a more advanced technique is needed to characterize phase coexistence in the 72 nm La-doped SSO, motivating our application of NISMA.

We employ PCA on the absorption images, which vary more strongly in energy than the reflectance images, to determine if there are correlations between contrasts visible across the spectroscopic regions measured. After identifying three statistically significant principal (linearly independent spectral) components in these data, a GMM roughly clusters our data into four distinct "phases" (details and visualization in *SI*). According to their spectral characteristics at low (Region I) and high energies (Regions II & III), we classify these as two structural phases, tetragonal (T) and orthorhombic (O), each subdivided into edge and bulk ("interior") phases, denoted $T_{edge}$, $T_{bulk}$, $O_{edge}$ and $O_{bulk}$, according to their respective spatial coordination. Fig. 5c maps these four phases according to pixels' highest probability of cluster membership within the GMM. The emergent striped pattern's orientation is consistent with predictions of the interface between the two structural phase interfaces (see *SI*). Linecuts of the confidence associated with phase assignment are shown in Fig. 5b (see *SI*), supporting the identification of edge and bulk phases, with $T_{edge}$ and $O_{edge}$ appearing closer to tetragonal and orthorhombic domain boundaries.

In the imaged region, the tetragonal phase is more prominent than the orthorhombic phase, and edge and bulk phases are about equally represented within structural phases (Fig. 5a). For each phase, we extract characteristic reflectance and absorption spectra (Fig. 5g-h), which are the average spectra of all pixels within a phase, normalized to the average spectrum. These characteristic spectra represent the differences in spectral behavior distinguishing these four phases. In Region I, the tetragonal phases have higher reflectance and lower absorption, which we attribute to the higher frequency phonon of the tetragonal phase compared to the orthorhombic phase. In Regions II and III, the $T_{edge}$ and $O_{edge}$ states present highest and lowest reflectance, respectively. To quantify the differences between the four phases, we fit these characteristic spectra as deviations from the Drude-Lorentz model parameters that describe the 72 nm La-doped

SSO film, as shown in Fig. 5g-h. To simplify the fitting, we model the film without the addition of electronic Lorentzian terms, like the model used in the analysis of the 19 nm tetragonal SSO of comparable carrier density. This simplification allows facile comparison of Drude fit parameters that quantify differences in mobility between these four phases. Further details on the fitting of characteristic spectra by a nonlinear least-squares method are given in *SI*. While the best-fit plasma frequency of the $T_{bulk}$ phase matches that for the area-average, the plasma frequency in the $O_{bulk}$ phase is lowered from 4480 cm$^{-1}$ to 4220 cm$^{-1}$. The two edge states showcase the most significant deviations in plasma frequency, increased to 4650 cm$^{-1}$ in the $T_{edge}$ phase and reduced to 4090 cm$^{-1}$ in the $O_{edge}$ phase. Fitted scattering rates were not different among these phases with any statistical significance (*SI* describes our certainty estimation). We conclude that differences in mobility between these coexisting phases arise primarily from variations in local effective mass (Fig. 5d). Indeed, $m^*$ calculated from the plasma frequency is about 29% higher (ranging from 19-40%) in $O_{edge}$ compared to $T_{edge}$, consistent with calculations reporting a 30% higher effective mass in the uniform orthorhombic phase relative to the tetragonal one[1]. However, our fits for bulk phases suggest a smaller difference in effective mass, with $O_{bulk}$ showcasing $m^*$ only 12% higher than $T_{bulk}$. (We note differences in $m^*$ as small as 2.8% or as large as 23% are consistent with variance in our data; see *SI*.) Our spatial identification of locally inhomogeneous plasma frequency highlights the nuanced impact of phase coexistence for achieving globally high mobility in SSO films.

To explain variations in optical response localized to the edges of structural domains, we investigate the accommodation strain emerging at domain boundaries. We use known lattice constants of the bulk structural phases as input to a Ginzburg-Landau free energy density in which orthorhombicity (quantified by an order parameter $\varphi(\vec{r})$) in SSO is frustrated by accompanying accommodation strain $\varepsilon(\vec{r})$ elastically deforming the substrate. Fig. 5f presents predictions by phase field modeling (details in *SI*) that striped equilibrium textures of $\varphi(\vec{r})$ minimize the system free energy across a broad temperature range in qualitative agreement with former findings[5,36] and imaging results (Fig. 5c). Fig. 5e superimposes the spatial dependence of simulated orthorhombicity $\varphi(\vec{r})$ with the accompanying accommodation strain $\varepsilon(\vec{r})$ (details in *SI*), revealing excess compressive and tensile strains exceeding 0.5% at tetragonal and orthorhombic domain edges that relaxes slowly towards domain interiors. We hypothesize these compressive and tensile edge strains inhomogeneously reduce (increase) the effective mass along tetragonal (orthorhombic) domain edges, a possible explanation for the optical response of "edge phases" $T_{edge}$ and $O_{edge}$ with spatial profiles highlighted by Fig. 5b. Thus, we attribute disparity in mobility of edge and bulk states to inhomogeneity of accommodation strain occurring at structural domain boundaries; this hypothesis remains to be confirmed through a direct spatially resolved probe of lattice structure, like TEM or nanoscale XRD[37].

We have introduced NISMA as a robust analysis method for nanoscale phase coexistence. By application of NISMA to a thin film of La-doped SSO, we categorized coexisting phases according to their spectroscopic structural and electronic responses. This robust analysis revealed that inhomogeneity in electronic properties in this material, insufficiently explained by a nonlocal plasmonic response, occurs even *within* domains of coexisting tetragonal and orthorhombic structural phases. Such intra-domain inhomogeneity should have significant ramifications for area-averaged mobilities of strain-engineered wide-bandgap semiconductors. Future work should combine nano-optical probes with correlative direct probes of inhomogeneous strain in these films through TEM[38] or nanoscale XRD[37]. The present work nevertheless showcases the power of unsupervised machine learning like NISMA to robustly analyze nanometer-resolved broadband hyperspectral imaging data and to identify inhomogeneous phases in optoelectronic materials.

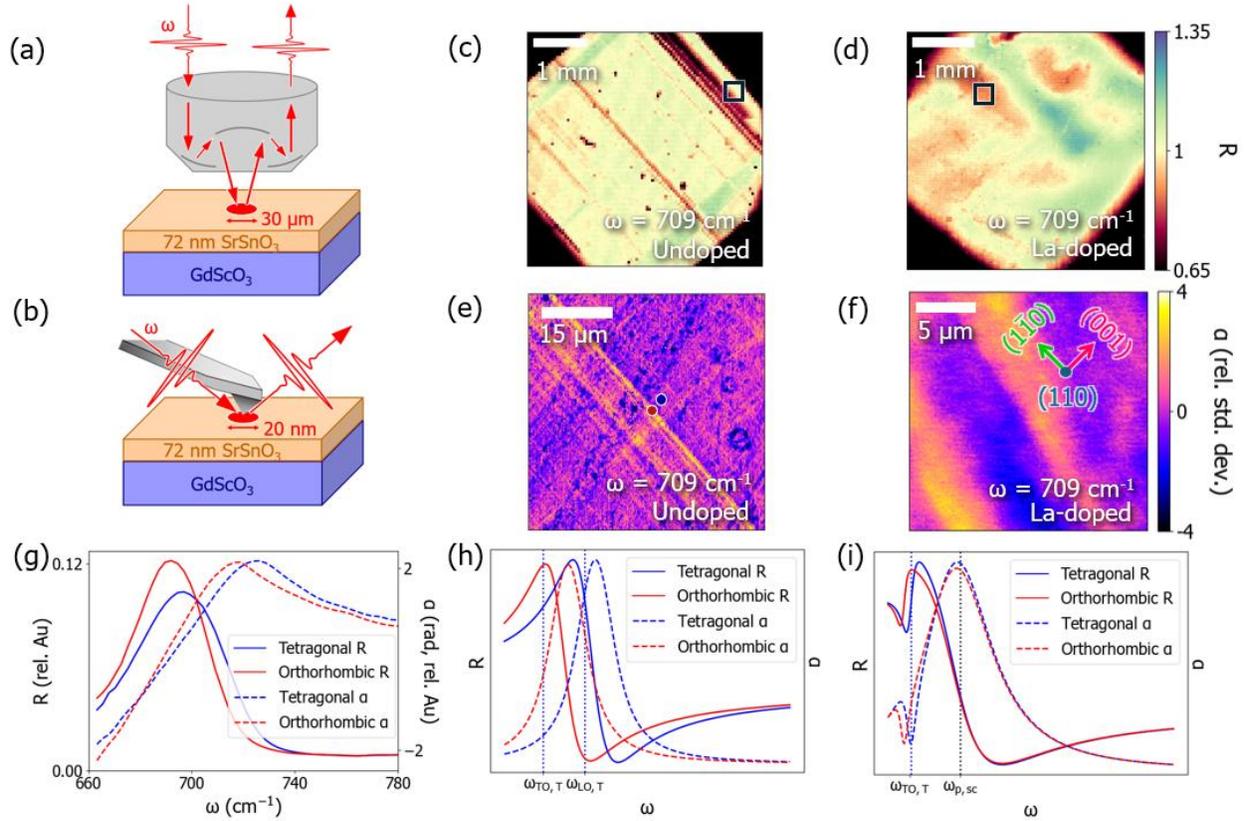

**Figure 1**. Comparison of undoped and La-doped SSO. (a) Schematic of a far-field, diffraction-limited measurement using a Schwarzschild objective. (b) Schematic of a near-field measurement using an AFM tip. (c-f) Far-field reflectance and near-field absorption imaging of undoped and La-doped SSO, using incident laser frequency of 709 cm$^{-1}$ for the far-field reflectance images (c-d) and near-field absorption images (e-f). (g) Nano-FTIR reflectance (R) and absorption (α) spectra of the undoped SSO, taken at positions indicated by the markers in (e). (h) Schematic of orthorhombic and tetragonal reflectance and absorption spectra. The vertical dotted lines indicate the transverse optical (TO) and longitudinal optical (LO) phonon frequencies for the tetragonal phase. (i) Simulated reflectance and absorption spectra for tetragonal and orthorhombic phases in
13

the doped film, assuming the same Drude response for both phases but different phonon frequencies.

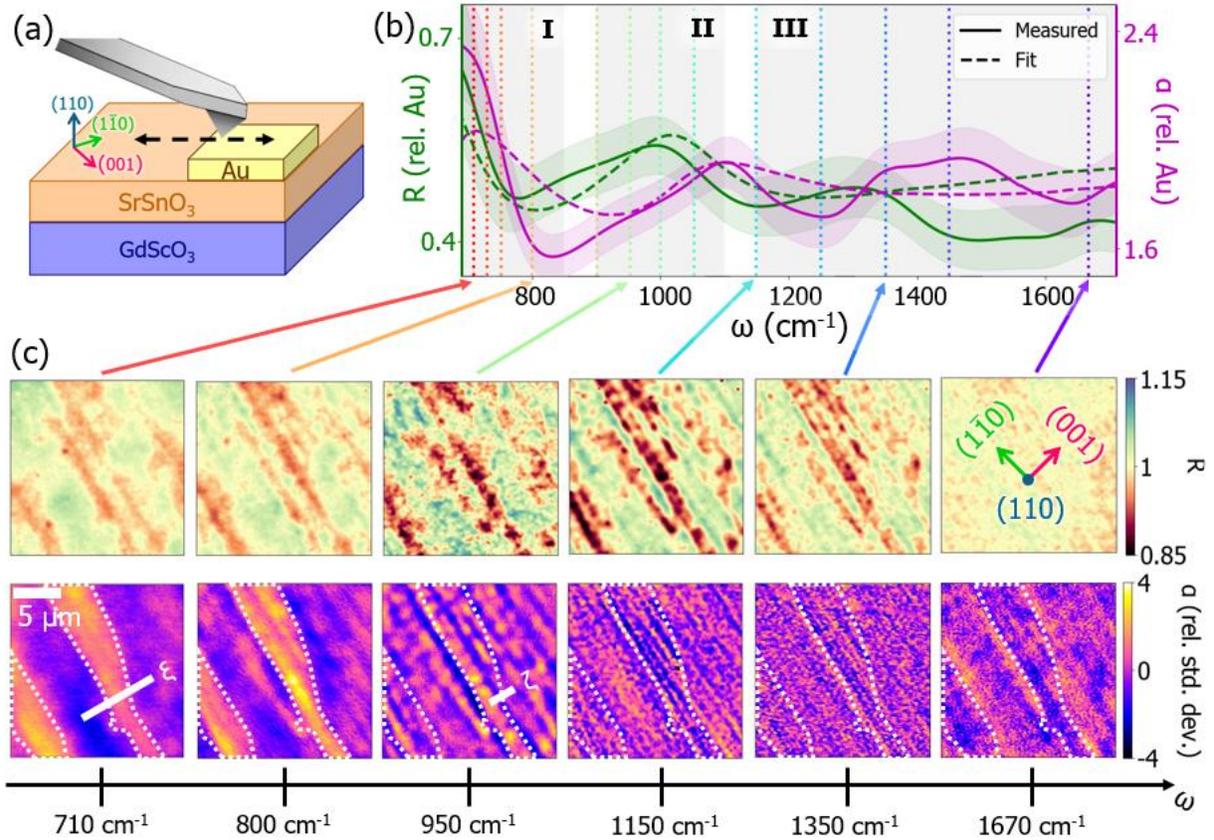

**Figure 2.** Nanoscale spectro-microscopy of La-doped SSO. (a) Schematic of near-field measurement using deposited gold pad for normalization. (b) Area-averaged near-field spectrum of the La-doped SSO. Dashed curves represent fit by the Drude-Lorentz model. Dotted vertical lines indicate all frequencies at which near-field images were taken. Shaded regions I, II and III indicate where the phonon (Region I) and electronic Lorentzian (Region II) and plasma response (Region III) have greater spectral weight. (c) Near-field reflectance (R) and absorption ($\alpha$) images of the La-doped SSO at selected frequencies. Scalebars for $\xi$ and $\zeta$ indicate lengths given in Fig. 3. Dashed lines indicate structural domains identified in Fig. 5c.

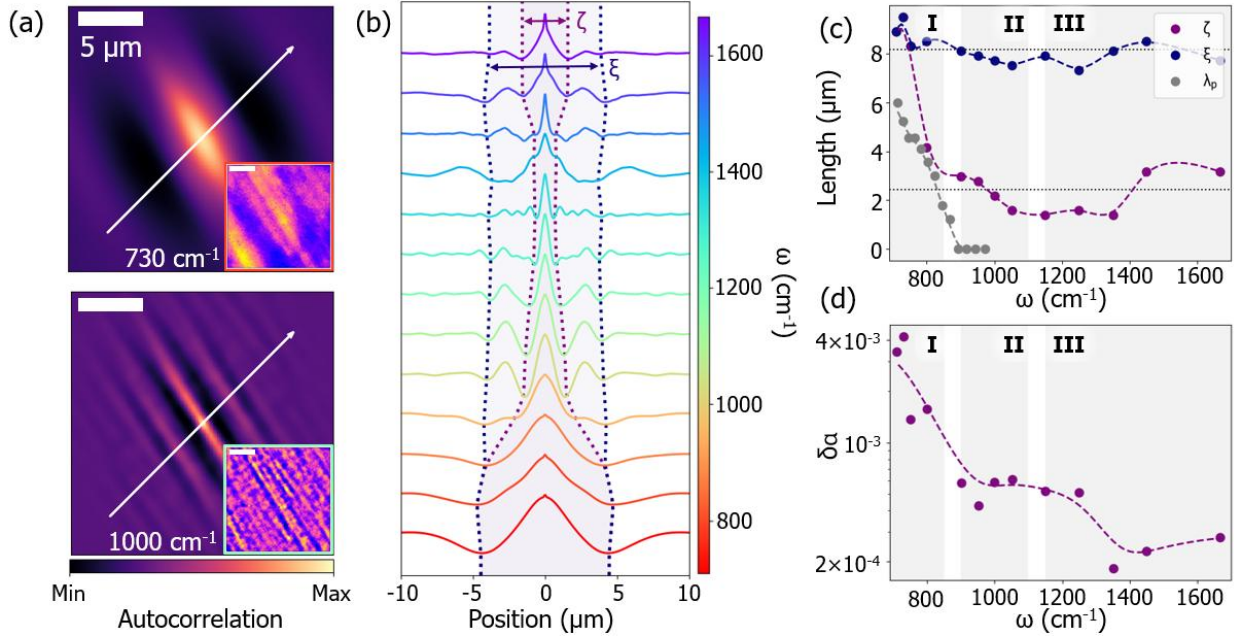

**Figure 3**. Absorption periodic length scales. (a) Autocorrelation maps for selected absorption images, shown as insets. (b) Linecuts of autocorrelation maps for each image, taken along arrows shown in (a). Shaded regions given by positions of the first minima and outer minima in linecuts. (c) Comparison of periodic length and plasmon wavelength as a function of frequency. The periodic lengths ξ and ζ were extracted from the linecuts in figure (b) as the distances between the indicated minima. The plasmon wavelengths were calculated from the distances between absorption extrema in Fig. 3b and rescaled by a factor of 72/19 to account for the different film thicknesses. (d) Variance in absorption (δα) as a function of frequency for the 72 nm La-doped SSO.

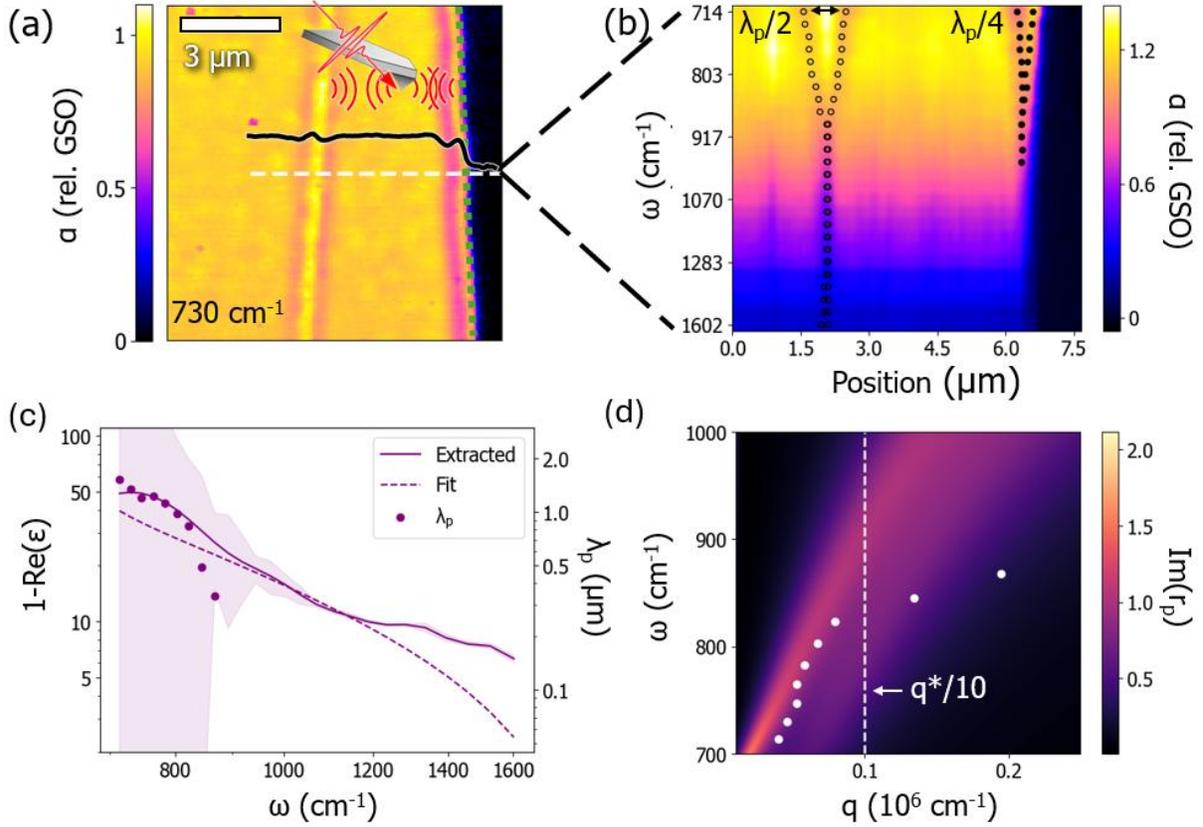

**Figure 4:** Plasmons in 19 nm La-doped SSO. (a) Absorption image at 730 cm$^{-1}$ of the etched boundary (indicated by the green dashed line) between the La-doped SSO (left of the boundary) and the GSO substrate (right of the boundary), with superimposed absorption linecut taken along the dotted white line. (b) Absorption linecuts taken along the dotted white line in (a), as a function of frequency. The distances between absorption extrema around the sample-substrate boundary and a scratch on the sample surface indicate the plasmon wavelength, $\lambda_p$. (c) Frequency dependent permittivity in the 19 nm La-doped SSO extracted from data (solid curve), fit using the Drude-Lorentz model (dashed curve), and calculated from the plasmon wavelengths $\lambda_p$ measured around the scratch in the sample (datapoints). (d) Plasmon dispersion calculated from the Drude-Lorentz fit shown in (c). Datapoints in white are calculated from the plasmon wavelengths measured

around the scratch in the sample. The white dashed line indicates one tenth of the typical momentum $q^* = 1/a$ supplied by the tip of radius $a \approx 10\ nm$.

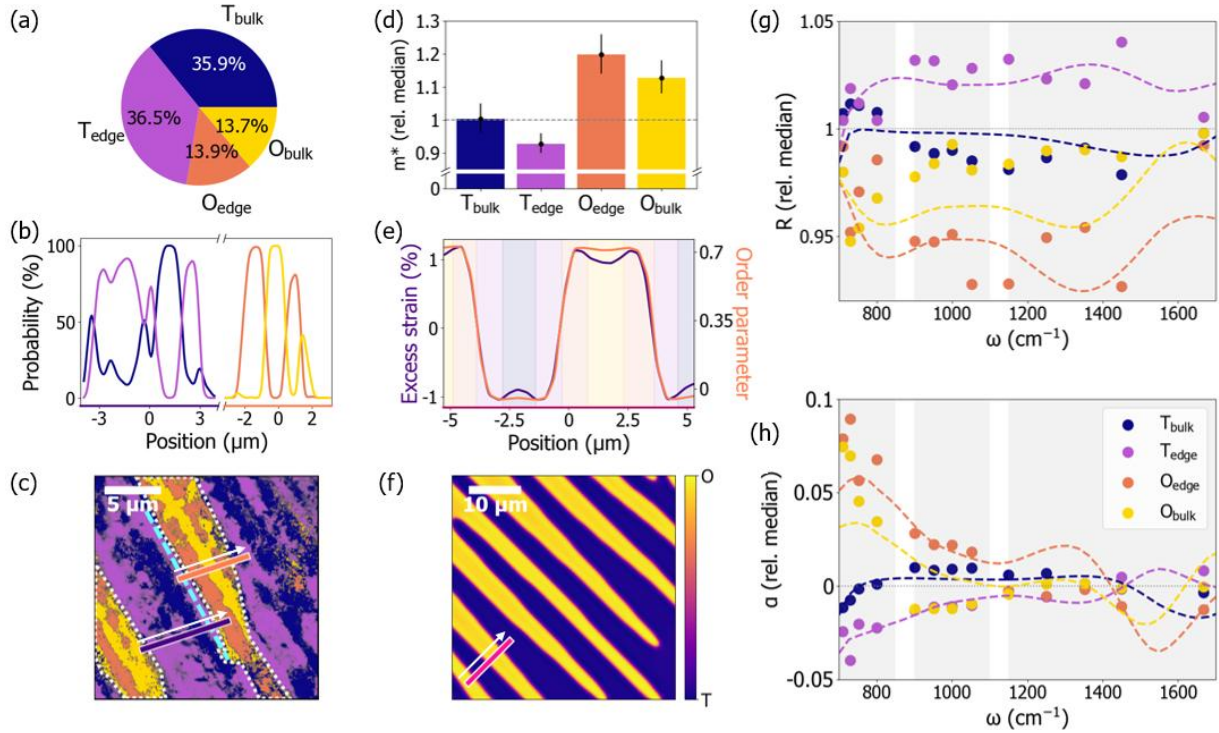

**Figure 5.** Principal component analysis of La-doped SSO. (a) Proportion of image area represented by each identified phase. (b) Linecuts of assignment probability for the tetragonal (T) and orthorhombic (O) phases. Linecuts are taken along the indigo (for tetragonal phases) and orange (for orthorhombic phases) lines shown in (c). (c) Spatial assignment of four identified phases in La-doped SSO. The blue dashed line shows the theoretically calculated interface between the two structural phases. The dashed lines indicate the identified structural domains. (d) Bar chart of the effective mass calculated for each phase, relative to that for the average spectrum. (e) Linecut taken along the pink line in (f) of strain field simulation. Shading indicates approximate locations of bulk and edge tetragonal and orthorhombic states. (f) Simulation of coexisting orthorhombic and tetragonal stripes in SSO. (g-h) Characteristic reflectance and absorption spectra for each phase, normalized to the average spectrum, with fits shown as dashed lines.

ASSOCIATED CONTENT

**Supporting Information**.

AUTHOR INFORMATION

**Corresponding Author**

**Present Addresses**

†Max-Planck-Institute for Chemical Physics of Solids, 01187 Dresden, Germany

**Author Contributions**

**Funding Sources**

**Notes**

ACKNOWLEDGMENT

# Supplementary Information

# Machine Learning-Assisted Nano-imaging and Spectroscopy of Phase coexistence in a Wide-Bandgap Semiconductor

*Alyssa Bragg[1], Fengdeng Liu[2], Zhifei Yang[1,2], Nitzan Hirshberg[1], Madison Garber[1], Brayden Lukaskawcez[1], Liam Thompson[1], Shane MacDonald[1], Hayden Binger[1, †], Devon Uram[1], Ashley Bucsek[3], Bharat Jalan[2], Alexander McLeod[1]*

[1]School of Physics and Astronomy, University of Minnesota – Twin Cities, Minneapolis, Minnesota 55455, United States

[2]Department of Chemical Engineering and Materials Science, University of Minnesota – Twin Cities, Minneapolis, Minnesota 55455, United States

[3]Department of Mechanical Engineering, University of Michigan, Ann Arbor, Michigan 48109, United States

†Max-Planck-Institute for Chemical Physics of Solids, 01187 Dresden, Germany

**Far-field infrared microscopy**

Far-field infrared microscopy was performed using the Phase Resolved Infrared Spectroscopy Microscope (PRISM), a custom device designed to measure reflectance amplitude and phase of infrared light while scanning the illuminating laser across the sample. The setup involves an asymmetric Michelson-based interferometer to allow for the extraction of both amplitude and phase of the reflected signal using a pseudoheterodyne technique[1], in which the optical path length difference between the reference and sample beams is sinusoidally modulated. Mid-infrared light was sourced from a difference-frequency generation module pumped by the fiber feedback-stabilized optical parametric amplifier of an ultrafast laser system (MIR, AlphaHP, and Primus units from Stuttgart Instruments). The PRISM scan head holds the Thorlabs LMM25XF-P01 Schwarzschild objective lens which focuses the sample beam onto the sample, as well as a flat gold mirror to reflect the reference beam back to the cryogenically cooled mercury cadmium telluride photoconductive detector (Judson Teledyne). Thorlabs LNR502 linear translation stages are used to move the scan head to scan the laser across the surface of the sample.

**Scanning near-field optical microscopy measurements**

We conducted nanoscale infrared imaging and spectroscopy of SSO using a Park Systems NX-10 atomic force microscope (AFM) modified to function as a scattering-type near-field optical microscope (s-SNOM), using a home-built asymmetric Michelson interferometer with dual imaging and spectroscopy modalities. Mid-infrared light was sourced from a difference-frequency generation module pumped by the fiber feedback-stabilized optical parametric amplifier of an ultrafast laser system (MIR, AlphaHP, and Primus units from Stuttgart Instruments), supplying 500 ps infrared pulses with 50 $cm^{-1}$ bandwidth at 40 MHz repetition rate. We focused this light to the sharp tip of a metallic microcantilevered AFM probe, which scattered in a manner determined by the complex permittivity of the sample volume proximate to the tip, whose apex radius determines the optical spatial resolution. A cryogenically cooled mercury cadmium telluride photoconductive detector (Judson Teledyne) recorded scattered infrared light which we demodulated (HF2LI from Zurich Instruments) at the second harmonic of the AFM probe oscillation (at 75 KHz, using an FM-PtSi probe from NanoSensors) for background suppression. By placing the s-SNOM and sample at one arm of the interferometer we could simultaneously extract the amplitude and phase of scattered light[1], which we associate with the coefficients of reflectance (R) and absorption (α), respectively. This latter association[2] follows from considering that the scattered field is proportional to the probe-sample polarization, whereas absorption arises when its time derivative shifts into phase with the incident field. We conducted the nano-infrared imaging and spectroscopy measurements on the SSO samples via pseudo-heterodyne imaging[3] and nanoscale Fourier transform infrared spectroscopy methods[2], respectively. NanoFTIR spectra were synthesized by merging several distinct spectra acquired at incremental tunings of the characteristic energy of the MIR stage output (from 600 to 2000 $cm^{-1}$).

**Film growth**

The La-doped $SrSnO_3$ thin films were grown using hybrid MBE. This approach employs conventional effusion cells for lanthanum and strontium, hexamethylditin (HMDT) as a metal-organic precursor for tin, and an inductively coupled radio frequency (RF) plasma for oxygen. The reader is referred to Prakash et al. for more details about the technique as applied to $BaSnO_3$[4]. All films were grown at a fixed substrate temperature of 950°C as measured with a floating thermocouple. The substrates were cleaned in situ with oxygen plasma for 25 minutes prior to film deposition. Sr was sublimed from a titanium crucible with its beam equivalent pressure (BEP) fixed at 6×10$^{-8}$ mbar as measured by a retractable beam flux monitor before growth. The hexamethylditin (HMDT, by which Sn was supplied) vapor entered the chamber through a heated gas injector (E-Science, Inc.) in an effusion cell port that was in direct line-of-site to the substrate, and the BEP ratio of Sn to Sr was kept at 550 to ensure stoichiometric growth. The oxygen flow was set to 0.7 standard cubic centimeters per minute (sccm) to achieve an oxygen background pressure of 5×10$^{-6}$ Torr while applying 250 watts of RF power to the plasma coil. The La cell temperature was kept at 1190°C to provide carriers for La-doped SrSnO$_3$. These conditions achieved a growth rate of ~33 nm per hour, and each film was grown for 130 minutes.

**High resolution X-ray diffraction (HRXRD)**

We grew ~72 nm SrSnO$_3$/GdScO$_3$ (110) and La-doped SrSnO$_3$/GdScO$_3$ (110) thin films for comparison to investigate the structural properties of SrSnO$_3$ exhibiting two phases using hybrid molecular beam epitaxy (MBE). Figure S1(a) shows the high resolution X-ray diffraction (HRXRD) 2$\theta$-$\omega$ coupled scan of SrSnO$_3$ (001)$_{pc}$/GdScO$_3$ (110) thin film, which exhibits two distinct peaks and clear Laue oscillations. We extracted the out-of-plane lattice parameters of two peaks to be 4.114 Å (corresponding to the tetragonal phase) and 4.094 Å (corresponding to the orthorhombic phase[5,6]). We also quantitively determined the area ratio of two phases ($A_\text{tetra}$/$A_\text{orth}$) to be 1.6 through Gaussian fitting, which indicates more tetragonal phase in the film. Figure S2(b) shows the HRXRD 2$\theta$-$\omega$ coupled scan of La:SrSnO$_3$ (001)$_{pc}$/ GdScO$_3$ (110) thin film, which shows a coalescence of two peaks. We extracted the out-of-plane lattice parameters of two peaks to be 4.105 Å (corresponding to the tetragonal phase) and 4.076 Å (corresponding to the orthorhombic phase[5,6]) and found $A_\text{tetra}$/$A_\text{orth}$ = 0.08 through Gaussian fitting, which suggests more orthorhombic phase in the film.

**Phonons as Lorentz oscillators**

Infrared spectra are acquired when photons from incident infrared light couple to dipoles formed by optical phonon modes of the lattice unit cell. For a three-dimensional unit cell with N atoms there are generally 3 phonon acoustic branches and 3N-3 phonon optical branches[7]. The optical branches further breakdown into N-1 longitudinal (LO) modes and 2N-2 transverse (TO) modes[7]. In LO modes the atoms are displaced in the direction of the propagation of the wave, and in TO modes the atoms are displaced perpendicular to the wave.

Permittivity is a complex quantity describing the interaction of a material with an electric field. It is a useful quantity for studying optical phonons as the real part, $\epsilon_1$, measures the energy stored through polarization, and the imaginary part, $\epsilon_2$, measures the absorption of incident light. For $n$ IR-active phonon modes, the permittivity as a function of incident light frequency, $\varepsilon(\omega)$, is described by

$$\epsilon(\omega) = \epsilon_\infty + \sum_n \frac{A_n \omega_{n,TO}^2}{\omega_{n,TO}^2 - \omega^2 - i\omega\gamma_{n,TO}} \qquad S1$$

where $A = \frac{\omega_{LO}^2 - \omega_{TO}^2}{\omega_{TO}^2}$, $\omega_{LO}$ is the frequency of the longitudinal optical mode, $\omega_{TO}$ is the frequency of the transverse optical mode, $\omega$ is the frequency of incident light, $\varepsilon_\infty$ is the permittivity in the high frequency limit, and $\gamma_{TO}$ is the phonon damping[8]. The permittivity can also be related to the conductivity, $\sigma$, by

$$\epsilon(\omega) = \epsilon_\infty + \frac{4\pi i\sigma}{\omega t} \qquad S2$$

where $\sigma$ is the 2D conductivity of the film and $t$ is the thickness of the film[8]. Then, the reflectivity as a function of permittivity is given by

$$R^* = \frac{\sqrt{\epsilon} - 1}{\sqrt{\epsilon} + 1} = R(\omega)e^{i\phi(\omega)}. \qquad S3$$

**Tetragonal and orthorhombic phonons in undoped SSO**

The measured reflectance and absorption spectra for the tetragonal and orthorhombic phases in the 72 nm film of undoped SSO are modelled using the permittivity given by Equation S1. The comparison of the fit to the data are shown in Fig S2, and the Lorentz parameters used to fit the spectra are given in Table S1.

|  | $\varepsilon_\infty$ | A | $\omega_{TO}$ (cm$^{-1}$) | $\gamma_{TO}$ (cm$^{-1}$) |
|---|---|---|---|---|
| Orthorhombic | 4.6 | 1.12 | 654 | 25 |
| Tetragonal | 4.6 | 0.91 | 670 | 36 |

**Table S1:** Fit phonon parameters for the 72 nm thick undoped SSO investigated in this work.

**The Drude-Lorentz model**

The electronic response of a doped semiconductor is dominated by the plasma oscillation at the screened plasma frequency $\omega_{p,sc} = \sqrt{\frac{4\pi n e^2}{m^* \epsilon_\infty}}$, where $n$ is the carrier concentration, $m^*$ is the effective mass of carriers and $\epsilon_\infty$ is the permittivity in the high frequency limit. Lorentz oscillators approximate the response of optical phonons such that when one optical phonon dominates, the permittivity in the Drude-Lorentz[9] model is given simply by

$$\epsilon(\omega) = \epsilon_\infty - \frac{\omega_p^2}{\omega(\omega + i\gamma_p)} + \frac{A_{TO}\omega_{TO}^2}{\omega^2 - \omega_{TO}^2 - i\omega\gamma_{TO}}. \qquad S4$$

Here $\gamma_p$ is the electron scattering rate, $A_{TO} = \frac{\omega_{LO}^2 - \omega_{TO}^2}{\omega_{TO}^2}$, $\omega_{LO}$ and $\omega_{TO}$ are the frequencies of the longitudinal and transverse optical phonons, and $\gamma_{TO}$ is the optical phonon scattering rate. Since phonon and plasma responses of SSO are prominent at separate energies, nano-imaging in distinct energy regimes can disentangle structural and electronic properties of coexisting domains[10]. Fig

S3 shows hypothetical relative differences (normalized to median) in spectra arising from the two structural phases in La-doped SSO, assuming both phases have the same Drude response (hypothetical spectra shown in main text Fig 1i).

**Drude-Lorentz parameters of 72 nm La-doped SSO**

The near-field spectrum of the 72 nm La-doped SSO was fit to the Drude-Lorentz model with a Drude term, a Lorentzian phonon mode, and two additional Lorentz terms representing electronic oscillations, in the form

$$\epsilon(\omega) = \epsilon_\infty - \frac{\omega_p^2}{\omega(\omega + i\gamma_p)} + \frac{A_{TO}\omega_{TO}^2}{\omega^2 - \omega_{TO}^2 - i\omega\gamma_{TO}} + \frac{A_{el,1}\omega_{el,1}^2}{\omega^2 - \omega_{el,1}^2 - i\omega\gamma_{el,1}} + \frac{A_{el,2}\omega_{el,2}^2}{\omega^2 - \omega_{el,2}^2 - i\omega\gamma_{el,2}}. \quad S5$$

For the 72 nm La-doped SSO, the phonon parameters $\omega_{TO}$, $\gamma_{TO}$ and $A_{TO}$ were the average of the phonon parameters found for the tetragonal and orthorhombic phases in the undoped SSO (Table S1). The values of $\varepsilon_\infty$ and $\gamma_p$ were taken from the fit of the 19 nm La-doped SSO (Table S4). The parameters used for the electronic Lorentzian terms and the value of $\omega_p$ were found by least-squares fitting to the near-field spectrum of 72 nm La-doped SSO. The full set of parameters used is given in Tables S2 and S3.

| $\varepsilon_\infty$ | $\omega_p$ (cm$^{-1}$) | $\gamma_p$ (cm$^{-1}$) | $A_{TO}$ | $\omega_{TO}$ (cm$^{-1}$) | $\gamma_{TO}$ (cm$^{-1}$) |
|---|---|---|---|---|---|
| 7.36 ± 0.567 | 4480 ± 70.3 | 220 ± 28.7 | 1.015 ± 0.364 | 663.5 ± 13.8 | 30.5 ± 0.908 |

**Table S2:** Drude-Lorentz parameters for the Drude and phonon terms used to model the spectra for the 72 nm La-doped SSO used in this work.

| $A_{el,1}$ | $\omega_{el,1}$ (cm$^{-1}$) | $\gamma_{el,1}$ (cm$^{-1}$) | $A_{el,2}$ | $\omega_{el,2}$ (cm$^{-1}$) | $\gamma_{el,2}$ (cm$^{-1}$) |
|---|---|---|---|---|---|
| 4.37 ± 0.276 | 988 ± 8.8 | 116 ± 18.1 | 19 ± 0.43 | 1930 ± 23.6 | 1370 ± 48.3 |

**Table S3:** Drude-Lorentz parameters for the additional electronic Lorentzian terms used to model the spectra for the 72 nm La-doped SSO used in this work.

We interpret the presence of the first electronic Lorentzian term as a possible polaron response[11–13]. We interpret the presence of the second electronic Lorentzian terms as corrections to the simplified Drude model used here. The extended Drude model[14,15] may also be used to account for frequency dependence of Drude parameters, namely the scattering rate and effective mass. In this work, we do not account for frequency dependence of such parameters, and instead incorporate the correction term as a Lorentzian oscillator within the simplified Drude model. This allows us to maintain consistency with the model used for the 19 nm La-doped SSO while also considering the full spectral response of the 72 nm film.

Using the known carrier concentration ($1.17 \times 10^{20}$ cm$^{-3}$), the effective mass can be calculated from the plasma frequency. The plasma frequency of 4480 cm$^{-1}$ corresponds to an effective mass of about $0.52 m_e$, which is about 66% higher than that reported for the tetragonal phase and 30% higher than that reported for the orthorhombic phase in previous literature[5].

**Image registration**

A critical component of NISMA is to synchronize the optical images such that they are aligned in pixel-by-pixel registry. We performed image registration on the topography images acquired simultaneously with the reflectance and absorption images. An optimal set of transformations (rotation and translation) to align each topography image to a reference topography image was found and applied to the associated reflectance and absorption images. Two example topography images taken on different days are shown after registration in Fig S4. The image taken at 800 cm$^{-1}$ laser illumination (Fig S2a) has been rotated clockwise to align with the image taken at 1150 cm$^{-1}$ (Fig S2b). The black pixels around the edges of the image represent spatial regions where data was not collected at that energy. All images were then cropped to remove such pixels that were not present in all images. In this way, all pixels are spatially aligned for the entire set of images. While such pixel-by-pixel registration is seldom implemented, this allows us to account for the unique spectroscopic response of over 42,000 pixels in our application of NISMA.

**Tetragonal SSO**

We investigated a 19 nm thin film of La-doped SSO with a carrier concentration of 9.75×10$^{19}$ cm$^{-3}$, comparable to the 72 nm thick La-doped SSO with a carrier concentration of 1.17×10$^{20}$ cm$^{-3}$. The 19 nm film is fully tetragonal, as confirmed by the shows the HRXRD 2$\theta$-$\omega$ coupled scan (Fig S5) with an extracted out-of-plane lattice parameter of 4.11 Å. Images are shown in Fig S6 in which the plasmon polaritons appear as rings around circular defects on the sample surface. The permittivity was fit to the Drude-Lorentz model. The phonon parameters were taken from the tetragonal phonon parameters found in the 19 nm undoped SSO, and the Drude parameters and the value of $\varepsilon_\infty$ were found by least-squares fitting. The full set of parameters used to fit the permittivity of the tetragonal SSO are given in Table S4. The value of $\omega_p = 4730$ (cm$^{-1}$) corresponds to an effective mass of about 0.39$m_e$, about 25% higher than previously reported for the tetragonal phase[5].

| $\varepsilon_\infty$ | $\omega_p$ (cm$^{-1}$) | $\gamma_p$ (cm$^{-1}$) | $A_{TO}$ | $\omega_{TO}$ (cm$^{-1}$) | $\gamma_{TO}$ (cm$^{-1}$) |
|---|---|---|---|---|---|
| 7.36 | 4730 | 220 | 0.91 | 670 | 26 |

**Table S4:** Drude-Lorentz parameters used to fit the permittivity of the 19 nm La-doped SSO.

**Extraction of permittivity from near-field spectra**

The near-field reflectance and absorption spectra relative are simulated to determine the polariton wavevector $q_p$ of the sample as a 2D material[16]. The polariton wavevector is related to the optical conductivity $\sigma_{2D}$ of the sample atop a substrate with permittivity $\epsilon_{sub}(\omega)$ by

$$q_p(\omega) = \frac{i\omega}{2\pi\sigma_{2D}(\omega)}. \qquad S6$$

The calculations are performed using the dimensionless value $\bar{q}_p = q_p/q^*$, where $q^* \sim 1/a$ is the characteristic momentum of the tip with radius $a$. Relating the permittivity and the conductivity of the sample by $\epsilon(\omega) = 1 + \frac{4\pi i \sigma(\omega)}{\omega}$, the permittivity of the sample with thickness $d$ is then obtained by

$$\epsilon(\omega) = 1 - \frac{2a}{\bar{q}_p d}. \qquad S7$$

The permittivity of the GSO substrate was modelled using previously fit Drude-Lorentz parameters[17].

The measured and simulated spectra for the 19 nm La-doped SSO are given in Fig. S7. The simulated spectra shown in Fig S7c. corresponds to the extraction of permittivity shown in Fig. 4c of the main text.

**Plasmon-polaritons**

In a 2D material on a substrate with permittivity $\epsilon_{sub}(\omega)$, the plasmon wavevector $q_p$ is given by[18]

$$q_p(\omega) = \frac{i\omega\kappa}{2\pi\sigma_{2D}(\omega)}, \qquad S8$$

where $\kappa = (1 + \epsilon_{sub}(\omega))/2$ is the average permittivity of the substrate and air surrounding the film. The plasmon wavelength $\lambda_p = 2\pi/q_p$ is then related to the permittivity of the film with thickness *d* by

$$\lambda_p(\omega) = \frac{\pi d}{\kappa}\big(1 - Re(\epsilon)\big). \qquad S9$$

In our surface plasmon-polariton imaging s-SNOM experiments, the conductive AFM tip also acts as an optical antenna to both launch and detect plasmons at the surface of our sample (schematic in main text, Fig. 4a). In our case, when tip-scattered photons simultaneously supply energies $\omega = \omega_p$ and momenta (inverse confinement) $q = q_p$ matching those of a plasmon resonance at the sample surface, plasmons are emitted. Notably, the sharp probe naturally supplies a range of momenta $q \leq 1/a$, with a ≈ 10 nm the probe radius, suitable for "launching" nanoscale plasmons[18–20]. In our case, real-space characteristics of emitted plasmons are observable only after they reflect from an edge, scratch, or other boundary on the sample, and propagate back towards the AFM tip. Probe-scattered light associates with both the local surface response (e.g. the "emitted" plasmon) as well as the nonlocal response from edge-reflected plasmons. In imaging, constructive (destructive) interference of electric fields from the emitted and reflected plasmons results in periodic bright (dark) fringes in the probe-scattered absorption response near sample edges, scratches or other plasmon reflectors. The plasmon half-wavelength is conventionally measurable by recording the distance between bright (or dark) fringes as a function of the energy of incident photons. When only a single fringe is observable, we apply the following simple model, which provides a satisfactory simplification of a more complete model[16] of probe-sample near-field interaction: Representing the $E_z$-field amplitude scattered by the substrate with a phasor $\tilde{e}_0$, and that of emitted plasmon with a phasor $\tilde{e}_p$, the total field scattered by the sample at the probe positioned a distance x from the sample edge is approximately $\tilde{E}_z(x) \approx \tilde{e}_0 + \tilde{e}_p(1 - \sqrt{\frac{a}{a+x}}e^{2iq_p x})$, with $q_p = 2\pi/\lambda_p$ the (complex) plasmon wave-vector and $a$ the probe radius. The second term in parentheses denotes the reflected ("image") plasmon field whose negative sign enforces the requirement $\tilde{E}_z(x) - \tilde{e}_0 \approx 0$ associated with vanishing plasmon polarization charge when the probe is scanned beyond the sample. The spatial dependence simplifies that of the polariton Green function, proportional to the radially (r) outgoing Hankel function $-iH_0^{(1)}(q_p r)$[21]. In our experiments, normalization to the non-absorptive substrate response renders $\tilde{e}_0$ unity and the plasmon phasor $\tilde{e}_p$ entirely imaginary (absorptive). In this model, the absorption response from the sample edge is first maximized at $x \approx \lambda_p/4$, and first minimized at $x \approx \lambda_p/2$.

In our measurements around the scratch on the sample surface, the x-component of the electric field vanishes at the scratch and the condition $\tilde{E}_x(x) - \tilde{e}_0 \approx 0$ is enforced. This introduces a 90-degree phase shift in the electric field compared to the case of plasmons scattered from the sample edge, and the absorption response is instead first minimized at $x \approx \lambda_p/4$. We measure the distance between the minima on either side of the scratch as $\lambda_p/2$. The plasmon wavelengths measured by these two methods are shown in Fig. S8d. Our measurements of $\lambda_p$ versus probe energy $\omega$ allows to record the plasmon-dispersion through $q_p = 2\pi/\lambda_p(\omega)$ or, equivalently, its inverse relation $\omega_p(q)$. In this work, we use the plasmon wavelengths measured around the scratch on the sample, which agree with the results of the extraction of permittivity (Fig. 4c-d of main text). The locations of the etched edges are less certain due to residual mask material and roughness or damage that may have an unknown effect on the measurement of plasmon wavelength.

**Principal component analysis (PCA) and clustering**

Principal component analysis (PCA) is used to reduce the dimensionality of large datasets. This is accomplished by representing the data in terms of principal components, which are vectors along the maximal variation in the data[22]. To illustrate how the principal components are defined, we consider the two-dimensional case, e.g., a dataset comprised of absorption images taken at two different energies, $\omega_1$ and $\omega_2$. In the original dataset, each pixel is a datapoint that has two associated measurements, $\alpha$ at $\omega_1$ and $\alpha$ at $\omega_2$. Although PCA can be conducted using complex values, incorporating both the reflectance and absorption images, we used only the absorption images since the subsequent clustering better represented the visible domain contrasts and clustered the data with greater confidence compared to both the complex approach and a reflectance-only approach. As shown schematically in Fig. S8a, the datapoints (pixels) form a distribution in the space of their absorption values at the two energies, and the first principal component is oriented along the maximal variation in this distribution. The second principal component is oriented along the next greatest direction of maximal variation orthogonal to the first. For a higher dimensional data set, more principal components are found in the same way, such that all principal components are orthogonal and represent decreasing variation in the dataset. Since the proportion of variance explained by each principal component is decreasing the first few principal components are chosen to represent the data. In this work, the first three principal components were used because the explained variance by each principal component begins to level out at the fourth principal component (Fig. S8b). This choice of principal components reduces the dimensionality while preserving much of the variance.

Each pixel is then clustered, or grouped, by a gaussian mixture model[23] according to its location in principal component space, shown in Fig. S8c. In the gaussian mixture model, the distribution of pixels in principal component space is represented by a set of superimposed gaussian probability distributions. Each gaussian probability distribution represents one cluster. Each pixel is then sorted into the cluster for which it has the highest probability. We define the confidence of each pixel's cluster assignment as the ratio of its probability within its assigned cluster's distribution to the sum of its probabilities in all cluster distributions, indicating how well a pixel is represented by its assigned phase. The confidence of all pixels' cluster assignments for the 72 nm La-doped SSO are mapped in Fig. S8d.

**Fitting of characteristic spectra in La-doped SSO**

We analyzed the differences in characteristic spectra for the $T_{edge}$, $T_{bulk}$, $O_{edge}$, and $O_{bulk}$ as deviations in Drude-Lorentz parameters from the 72 nm La-doped SSO, with phonon and Drude parameters given in Table S2, excluding the additional Lorentz term. The parameters associated with the phonon were held fixed according to the measured phonon parameters in the undoped film (Table S1). The Drude parameters were determined by least-squares fitting of the characteristic spectra. These deviations from the parameters used in fitting the average spectrum are summarized in Table S5.

|  | $T_{bulk}$ | $T_{edge}$ | $O_{edge}$ | $O_{bulk}$ |
|---|---|---|---|---|
| $\delta\varepsilon_\infty$ | 0.00545 ± 0.679 | 0.559 ± 0.576 | -1.20 ± 0.698 | -0.878 ± 0.625 |
| $\delta\omega_p$ (cm$^{-1}$) | -5.88 ± 95.8 | 173 ± 81.3 | -387 ± 98.4 | -260 ± 88.1 |
| $\delta\gamma_p$ (cm$^{-1}$) | -6.67 ± 43.7 | -5.01 ± 37.0 | 12.3 ± 44.8 | 24.0 ± 40.1 |
| $\delta A_{TO}$ | -0.105 | -0.105 | 0.105 | 0.105 |
| $\delta\omega_{TO}$ (cm$^{-1}$) | 6.5 | 6.5 | -6.5 | -6.5 |
| $\delta\gamma_{TO}$ (cm$^{-1}$) | 5.5 | 5.5 | -5.5 | -5.5 |

**Table S5:** Drude-Lorentz parameters used to fit the four characteristic spectra of the 72 nm La-doped SSO, relative to the parameters used for the average fit (Table S2).

**Strain calculations of SSO microstructure**

The calculations presented here are based on the crystallographic theory of martensite[24]. Originally developed to describe phase transformations in steel (hence the austenite/martensite terminology), it is applicable for any reversible, diffusionless, solid-to-solid phase transformation. Following the nomenclature of this theory, we will refer to the low-symmetry crystallographic phase as the martensite phase and the high-symmetry crystallographic phase as the austenite phase.

The tetragonal-to-orthorhombic phase transformation (martensite = tetragonal, austenite = orthorhombic) has two martensite variants, which can be defined by the stretch components of the deformation gradient that takes the austenite lattice to the martensite lattice, $\mathbf{U}_i$ where $i = 1, 2$ corresponds to each of the two variants

$$\mathbf{U}_1 = \begin{bmatrix} \alpha & 0 & 0 \\ 0 & \beta & 0 \\ 0 & 0 & \gamma \end{bmatrix}, \mathbf{U}_2 = \begin{bmatrix} \beta & 0 & 0 \\ 0 & \alpha & 0 \\ 0 & 0 & \gamma \end{bmatrix} \qquad S10$$

where $\alpha = a_O/a_T$, $\beta = b_O/b_T$, $\gamma = c_O/c_T$ are the transformation stretches. Superscripts O and T refer to the orthogonal and tetragonal phases, respectively.

In the case of a thin film, the compatibility requirements are less stringent than in bulk materials, because deformation in the out-of-plane direction can be considered to be free. As a result, we can consider the possibility of a compatible interface between austenite and a single martensite variant. Because we are working on very thin films, this (normally two-dimensional) interface is reduced to a line that exists somewhere in the epitaxial plane (perpendicular to the out-of-plane growth direction). The condition for compatibility between the austenite martensite phases along this line is

$$(\mathbf{QU}_i - \mathbf{I})\hat{e} = 0, \qquad \hat{e} \cdot \hat{e}_3 = 0 \qquad S11$$

where $\hat{e}$ is the compatible interface line, $\hat{e}_3$ is the out-of-place direction, and $\mathbf{Q}$ just allows for the possibility of a rotation of the martensite variant with respect to the austenite lattice.

The following methodology for solving Equation S11 is published in [25] and also reprinted in Ch. 10 of [24]:

1. Calculate $A = U_I^T U_I - I$.

2. Equation S11 has a solution if and only if $\hat{e}_3 \cdot \text{cof } A\hat{e}_3 \leq 0$.

3. To find $\hat{e}_3$, choose any mutually perpendicular unit vectors $\hat{e}_1$ and $\hat{e}_2$ in the plane of the film so that $\{\hat{e}_1, \hat{e}_2, \hat{e}_3\}$ form an orthonormal basis. The solution to S10 is given by

$$\hat{e} = \alpha \hat{e}_1 + \beta \hat{e}_2 \qquad S10$$

where $\alpha$, $\beta$ simultaneously satisfy the equations $\alpha^2 + \beta^2 = 1$ and $\alpha^2 \hat{e}_1 \cdot A\hat{e}_1 + 2\alpha\beta \hat{e}_1 \cdot A\hat{e}_2 + \beta^2 \hat{e}_2 \cdot A\hat{e}_2 = 0$.

The most important input to the thin film compatibility theory are the stress-free lattice parameters, ideally measured at room temperature (RT). The material parameters for a SrSnO3 (SSO) film on a (110) GdScO3 (GSO) substrate are shown in Table S6.

|  | a (Å) | b (Å) | c (Å) |
| --- | --- | --- | --- |
| **GSO substrate at room temperature** | 5.45 | 5.75 | 7.93 |
| **SSO austenite (tetragonal) at 1173 K**[26] | 5.7551 | 5.7551 | 8.1612 |
| **SSO martensite (orthorhombic) at 573 K**[26] | 5.7179 | 5.7279 | 8.0869 |
| **SSO martensite (orthorhombic) at 293 K**[27] | 5.7082 | 5.7035 | 8.0659 |

**Table S6:** SSO and GSO material parameters from literature.

Applying the material parameters shown in Table S6, we yield interesting results. Using the thermal coefficient of $\alpha_T = 1.0212 \times 10^{-5}$ estimated from comparing the orthorhombic SSO lattice parameters at 573 K and 293 K in Table S6, there are two possible austenite-martensite (i.e., tetragonal-orthorhombic) interfaces. Specifically, there are four possible $\hat{e}$ vectors as shown in Fig. S9a that provide solutions to Equation S11. These solutions come in pairs, where the two solutions in each pair oriented 180° apart, so each pair falls along the same line. In other words, four solutions yield two possible interfaces. The two possible interface lines are superimposed on the experimental results in Fig. S9b. Notice that one interface (extending from top left to bottom right) is an almost perfect match, so we will take this to be our identified interface. This solution is extremely sensitive to lattice parameters, and a very small change would have yielded completely different results. For example, if our thermal coefficient was $0.97 \times 10^{-5}$ (instead of $1.0212 \times 10^{-5}$), the two possible interfaces would both be almost perfectly horizontal. If our thermal coefficient was $1.15 \times 10^{-5}$, the two possible interfaces would be almost perfectly vertical. If our thermal coefficient did not fall within $0.97 \times 10^{-5} \leq \alpha_T \leq 1.15 \times 10^{-5}$, then there would have been zero possible interfaces, so the calculations are very sensitive to small changes in lattice parameters, and a very small change in our input parameters would have yielded drastically different interfaces or no interfaces at all.

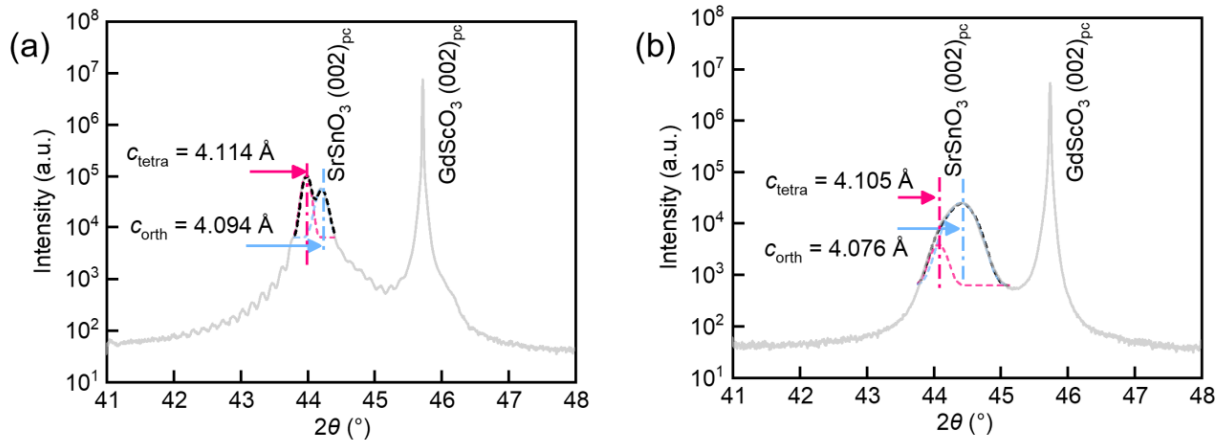

**Figure S1:** High resolution X-ray diffraction (HRXRD) 2$\theta$-$\omega$ coupled scans of (a) SrSnO$_3$ (001)$_{pc}$/ GdScO$_3$ (110) thin film and (b) La:SrSnO$_3$ (001)$_{pc}$/ GdScO$_3$ (110) thin film.

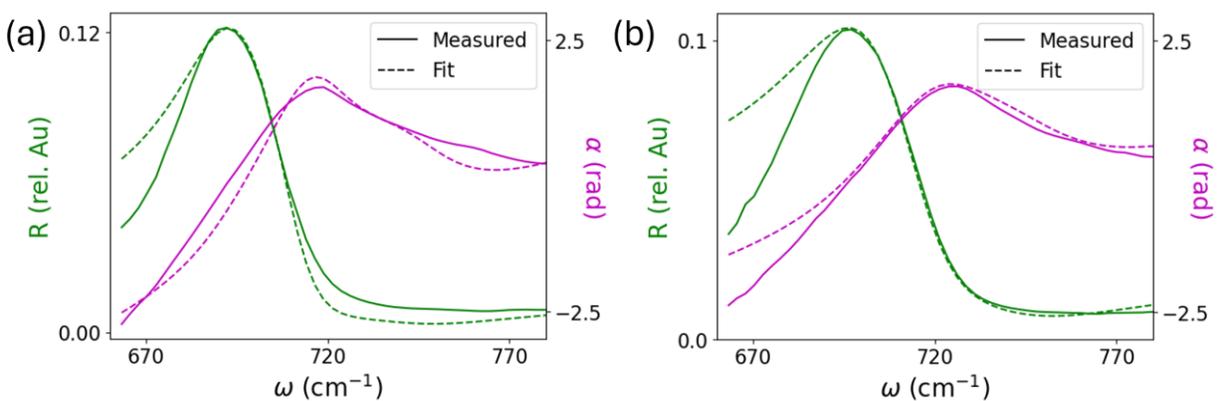

**Figure S2:** Spectra and fits for the (a) orthorhombic and (b) tetragonal phases in undoped SSO. The parameters used in fitting are given in Table S1.

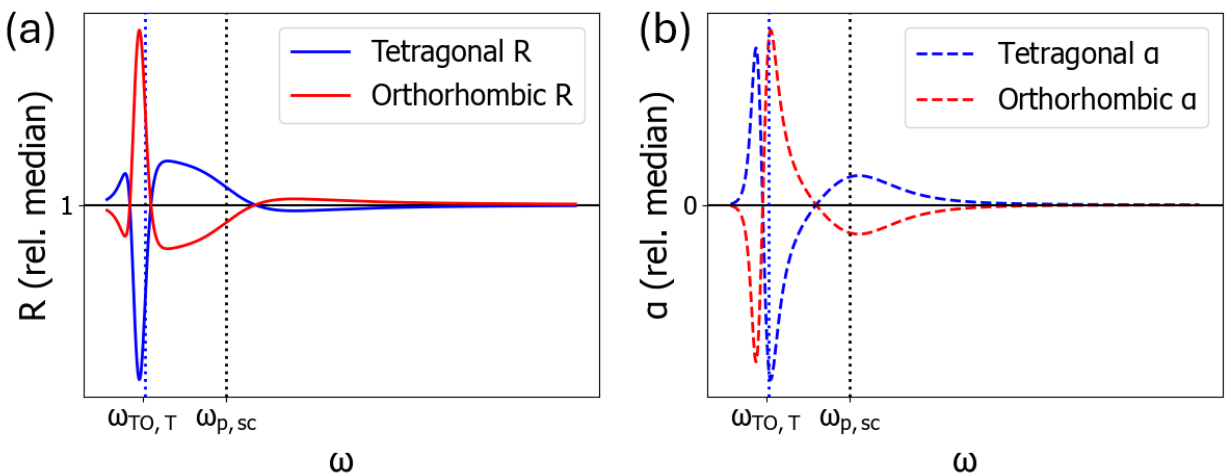

**Figure S3:** Simulated characteristic reflectance and absorption spectra for tetragonal and orthorhombic phases in the doped film, assuming the same Drude response for both phases but different phonon frequencies.

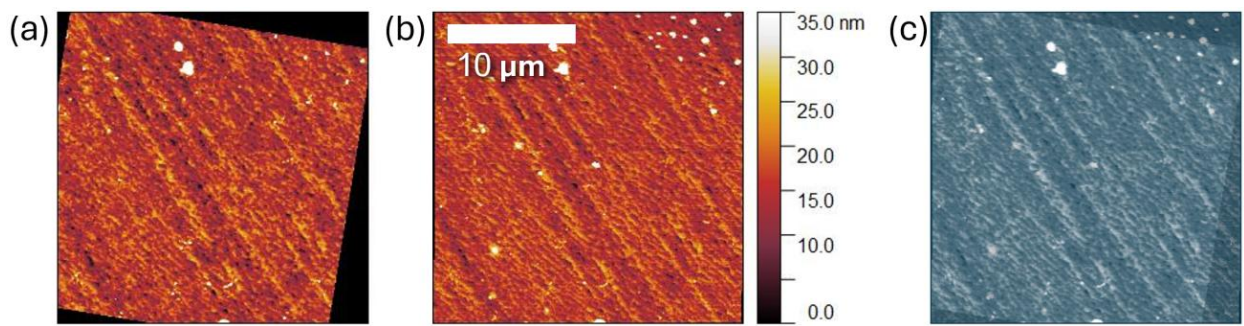

**Figure S4:** Registered topography images taken simultaneously with (a) 800 cm$^{-1}$ and (b) 1150 cm$^{-1}$ reflectance and absorption images. (c) Topography image in (a) overlayed at 50% transparency onto image in (b).

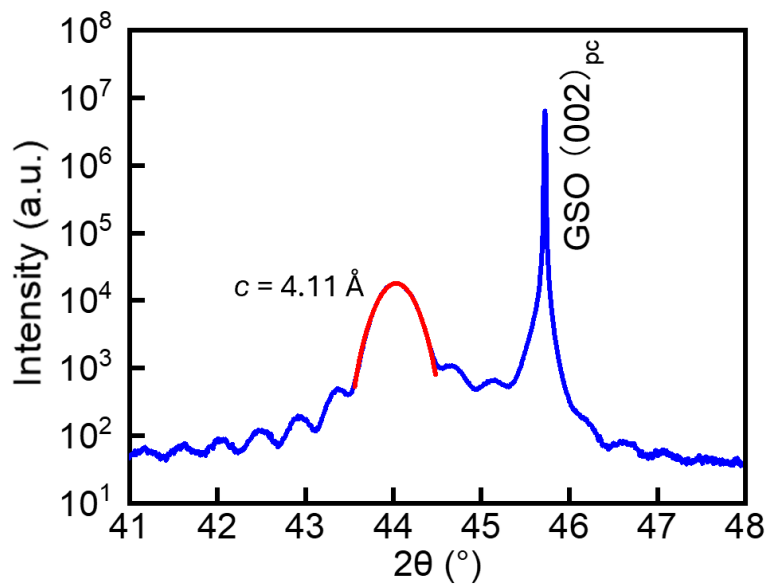

**Figure S5:** High resolution X-ray diffraction (HRXRD) $2\theta$-$\omega$ coupled scan of 19 nm SrSnO$_3$ (001)$_{pc}$/ GdScO$_3$ (110) thin film.

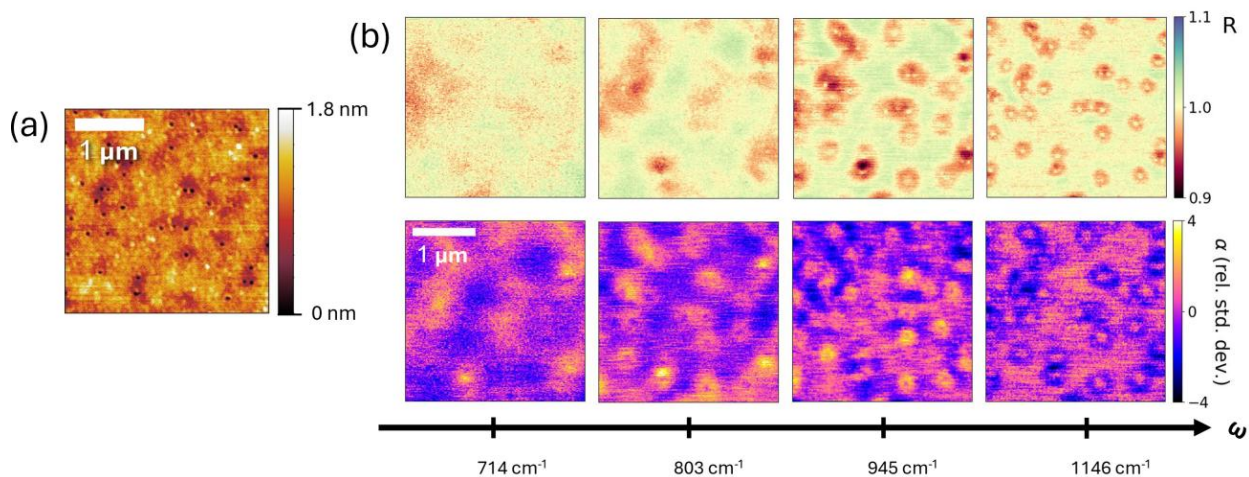

**Figure S6:** (a) Topography image of the 19 nm La-doped SSO. (b) Near-field reflectance (R) and absorption (α) images of the 19 nm La-doped SSO at selected frequencies, at the same region as shown in (a).

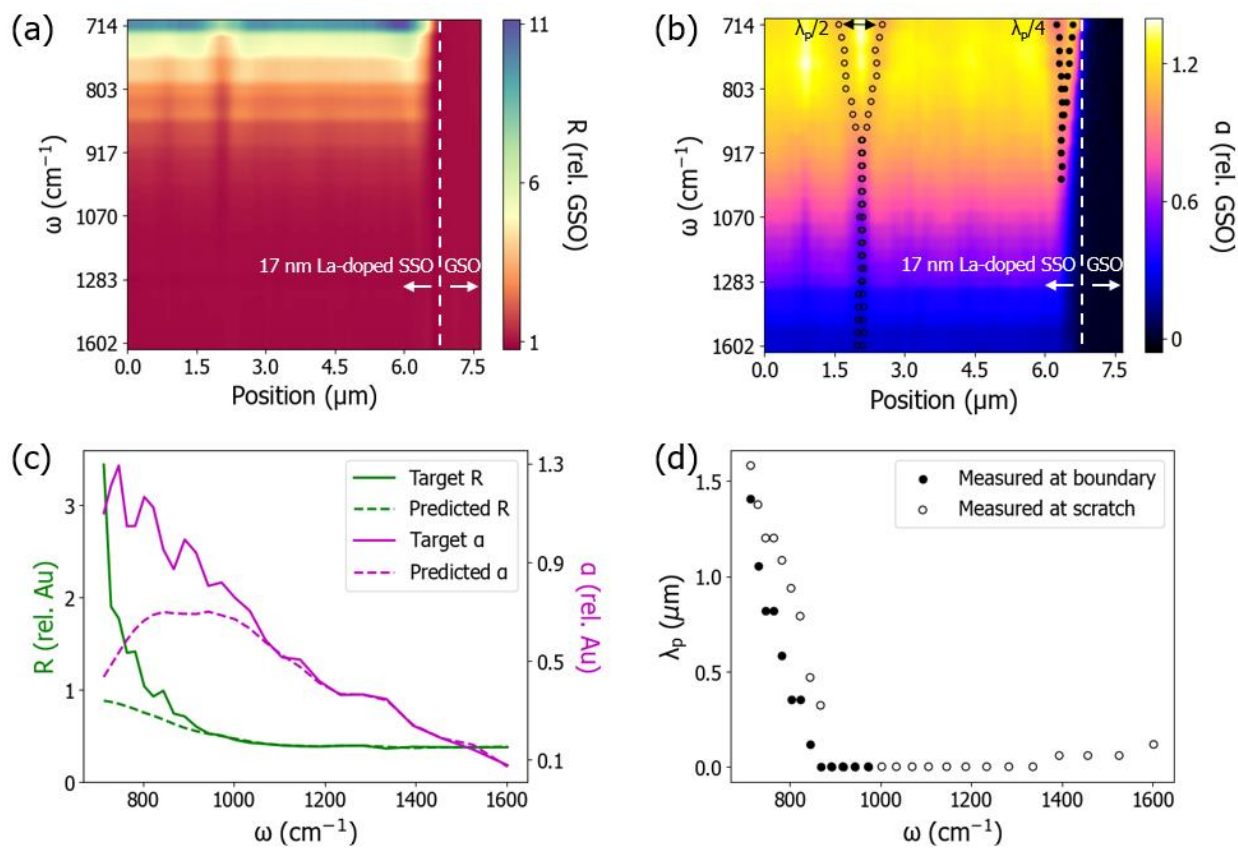

**Figure S7:** Permittivity extraction and plasmon wavelength for 19 nm La-doped SSO. Reflectance (a) and absorption (b) linecuts as a function of energy across the etched boundary between the sample and GSO substrate. (c) Prediction of spectra given by the extraction of permittivity for 19 nm La-doped SSO, fit to target spectra calculated from the reflection and absorption contrasts shown in (a) and (b). (d) Comparison of plasmon wavelengths measured at the boundary between the sample and the GSO substrate and at the scratch on the sample surface.

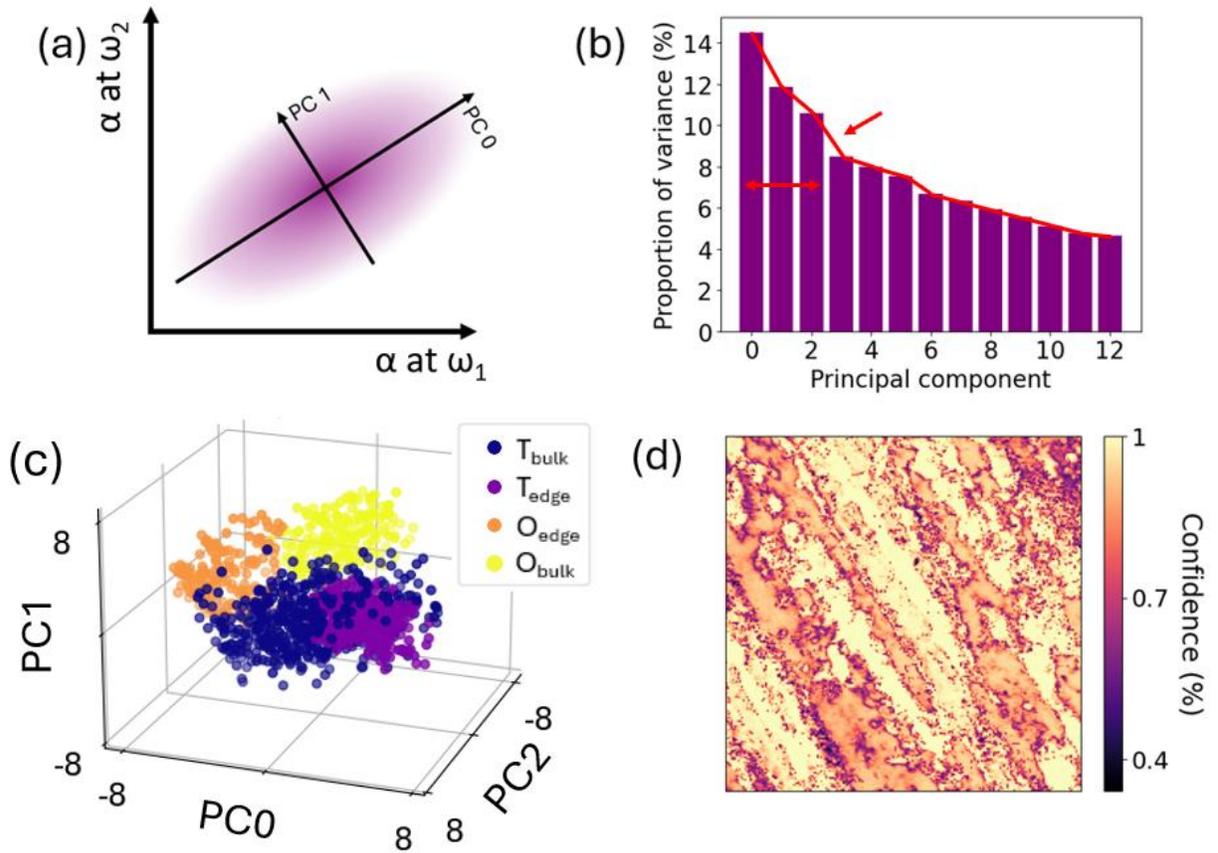

**Figure S8:** Principal component analysis of 72 nm La-doped SSO. (a) Schematic of principal components identified along variance in absorption images. (b) Proportion of variance in La-doped SSO images explained by each principal component. (c) Locations of pixels within each cluster in principal component space. (d) Map of confidence associated with cluster assignment.

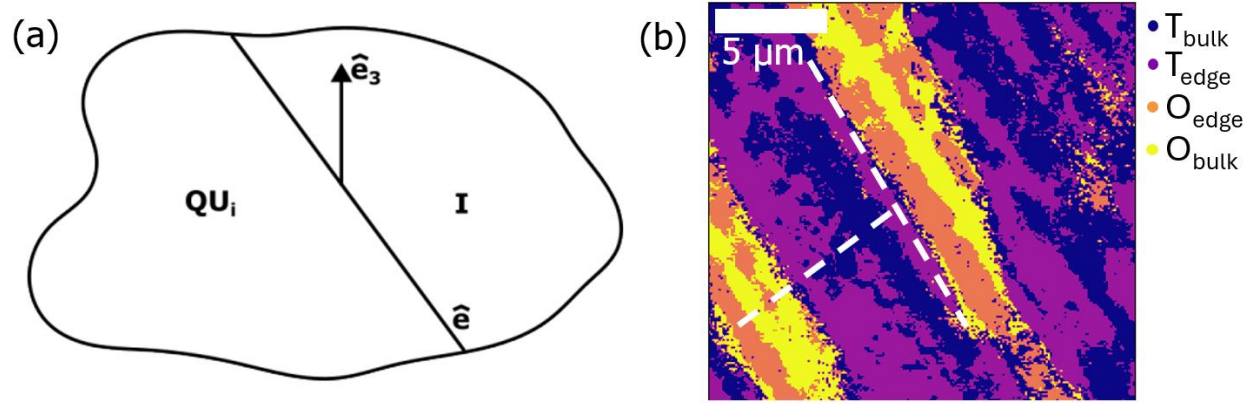

**Figure S9:** (a) Schematic of Equation S11. (b) Comparison of theoretical interface predicted by martensitic phase transformation and experimental results.